\documentclass[%
 reprint,
 nofootinbib,
 amsmath,amssymb,
 aps,
10pt,
floatfix]{revtex4-2}

\usepackage{graphicx}
\usepackage{lipsum}
\usepackage{hyperref}
\hypersetup{
    colorlinks=true,
    allcolors = {blue}
}
\usepackage[dvipsnames]{xcolor}
\usepackage{dcolumn}
\usepackage{bm}
\usepackage[mathlines]{lineno}

\usepackage{silence}
\usepackage{ulem}
\usepackage{color}
\usepackage{float}
\usepackage{bbold}
\usepackage{adjustbox}
\usepackage{pgfplots}
\usepackage{dirtytalk}
\usepackage{multirow}
\usepackage{isotope}
\usepackage[caption=false]{subfig}

\pgfplotsset{compat=1.18}

\usepackage{lineno}

\begin{document}
\preprint{APS/123-QED}

\title{{\Large Developments in NuWro Monte Carlo generator} \\~\\ {Contribution to the 25th International Workshop on Neutrinos from Accelerators }}

\author{Hemant Prasad}
\email{contact author hemant.prasad@uwr.edu.pl}
\author{Jan T. Sobczyk}%
\email{contact author jan.sobczyk@uwr.edu.pl}
 \affiliation{%
 Institute of Theoretical Physics\\
 University of Wroc{\l}aw, Plac Maxa Borna 9, 50-204 Wroc{\l}aw, Poland
}

\author{Artur M. Ankowski}
\affiliation{%
 Institute of Theoretical Physics\\
 University of Wroc{\l}aw, Plac Maxa Borna 9, 50-204 Wroc{\l}aw, Poland
}
\author{J. Luis Bonilla}
\affiliation{%
 Institute of Theoretical Physics\\
 University of Wroc{\l}aw, Plac Maxa Borna 9, 50-204 Wroc{\l}aw, Poland
}

\author{Rwik Dharmapal Banerjee}
\affiliation{%
 Institute of Theoretical Physics\\
 University of Wroc{\l}aw, Plac Maxa Borna 9, 50-204 Wroc{\l}aw, Poland
}

\author{Krzysztof M. Graczyk}
\affiliation{%
 Institute of Theoretical Physics\\
 University of Wroc{\l}aw, Plac Maxa Borna 9, 50-204 Wroc{\l}aw, Poland
}

\author{Beata E. Kowal}
\affiliation{%
 Institute of Theoretical Physics\\
 University of Wroc{\l}aw, Plac Maxa Borna 9, 50-204 Wroc{\l}aw, Poland
}



\begin{abstract}
In this article, we highlight physics improvements in the NuWro Monte Carlo event generator. The upcoming version of NuWro will incorporate the integration of the argon spectral function for quasi-elastic scattering, along with the MINER\ensuremath{\nu}A parametrization of the axial form factor. Additionally, the new release will feature the implementation of the Valencia 2020 model for meson exchange current. The previously used simplistic delta resonance model for single-pion production will be replaced by a more accurate Ghent hybrid model in the upcoming version of NuWro. We also discuss the recent advancements made by the Wroc{\l}aw Neutrino Group in applying machine-learning techniques to achieve model-independent reconstruction of lepton-nucleus interactions.
\end{abstract}

\maketitle


\section{\label{sec:Introduction}INTRODUCTION}
{\sc NuWro} is a Monte Carlo (MC) neutrino event generator developed at the University of Wroc{\l}aw since 2004~\cite{Juszczak:2005zs, Golan:2012wx}. It can operate in a broad energy spectrum ranging from a few hundred MeVs to hundreds of GeV, for both charged current (CC) and neutral current (NC) interactions in neutrino-nucleon and neutrino-nucleus scattering. Wroc{\l}aw Neutrino Group also provides a piece of software dedicated to electron-nucleus interaction called e{\sc Wro} which serves the purpose of validating models against electron scattering data. The code is identical to {\sc NuWro} in all aspects. Currently, eWro works only for CCQE interactions. In the future, eWro will be extended to other interaction channels as well.

The current version of {\sc NuWro} available on GitHub at \verb+https://github.com/nuwro/nuwro.git+ is 21.09.2. A new version is anticipated to be released in early 2025. 

\section{\label{sec:improvements}UPCOMING DEVELOPMENTS IN NuWro}
\subsection{\label{subsec:QE} Improvements in quasi-elastic channel}


The upcoming version of {\sc NuWro} will include the argon spectral function,
which was derived from a coincidence electron-scattering experiment conducted at the Jefferson Laboratory (JLab) Hall A~\cite{Banerjee:2023hub, PhysRevD.105.112002, PhysRevD.107.012005}. Charged lepton kinematics will be improved by accounting for the distortion caused by the nuclear Coulomb field~\cite{Banerjee:2023hub, PhysRevD.91.033005}. This improvement will be particularly valuable in the near future for the Short-Baseline Neutrino program at Fermilab, which includes experiments such as SBND~\cite{2024} and MiniBooNE~\cite{PhysRevD.64.112007}. 


The newly implemented spectral function approach with the MINER\ensuremath{\nu}A axial form factor was tested against the MicroBooNE CC$1p0\pi$ data \cite{PhysRevLett.125.201803}. The SF approach describes the data with the $\chi^2/\text{d.o.f}$ of 0.7, compared with $1.0$ for the LFG model, with the MINER\ensuremath{\nu}A axial form factor parametrization (see Table I of Ref.~\cite{Banerjee:2023hub}). The result indicates a consistency between the MINER\ensuremath{\nu}A and the MicroBooNE experiments.

\subsection{\label{subsec:MEC} {\sc NuWro} implementation of 2020 Valencia model}

{\sc NuWro} version 21.09.2 currently includes several inclusive models for MEC. These models are the Valencia model, the Marteau model, the SuSAv2 model, and the transverse enhancement (TE) model. 

The Valencia and SuSAv2 models are implemented by tabulating the five response functions, $W^{\mu\nu}$. This approach is both effective and efficient, as the same set of tables can be used to compute the inclusive cross section for any (anti)neutrino flavor at any energy. The individual components of \(W^{\mu\nu}\) describe the inclusive electroweak nuclear responses for a pair of correlated nucleons. However, $W^{\mu\nu}$ does not provide information regarding momenta of the outgoing nucleons or their isospins.

In MC generators, it is necessary to model the kinematics of the outgoing nucleon states explicitly. To accomplish this, the inclusive models are paired with their exclusive parts, which simulate the kinematics of the outgoing nucleon states. Up until now, the exclusive parts of all the MEC models are modeled identically using a simplistic phase space model proposed in Ref.~\cite{Sobczyk:2012ms}.

\begin{figure}[htbp!]
    \begin{minipage}[b]{0.7\linewidth}
        \begin{adjustbox}{width=\linewidth}
            \includegraphics[width=\linewidth]{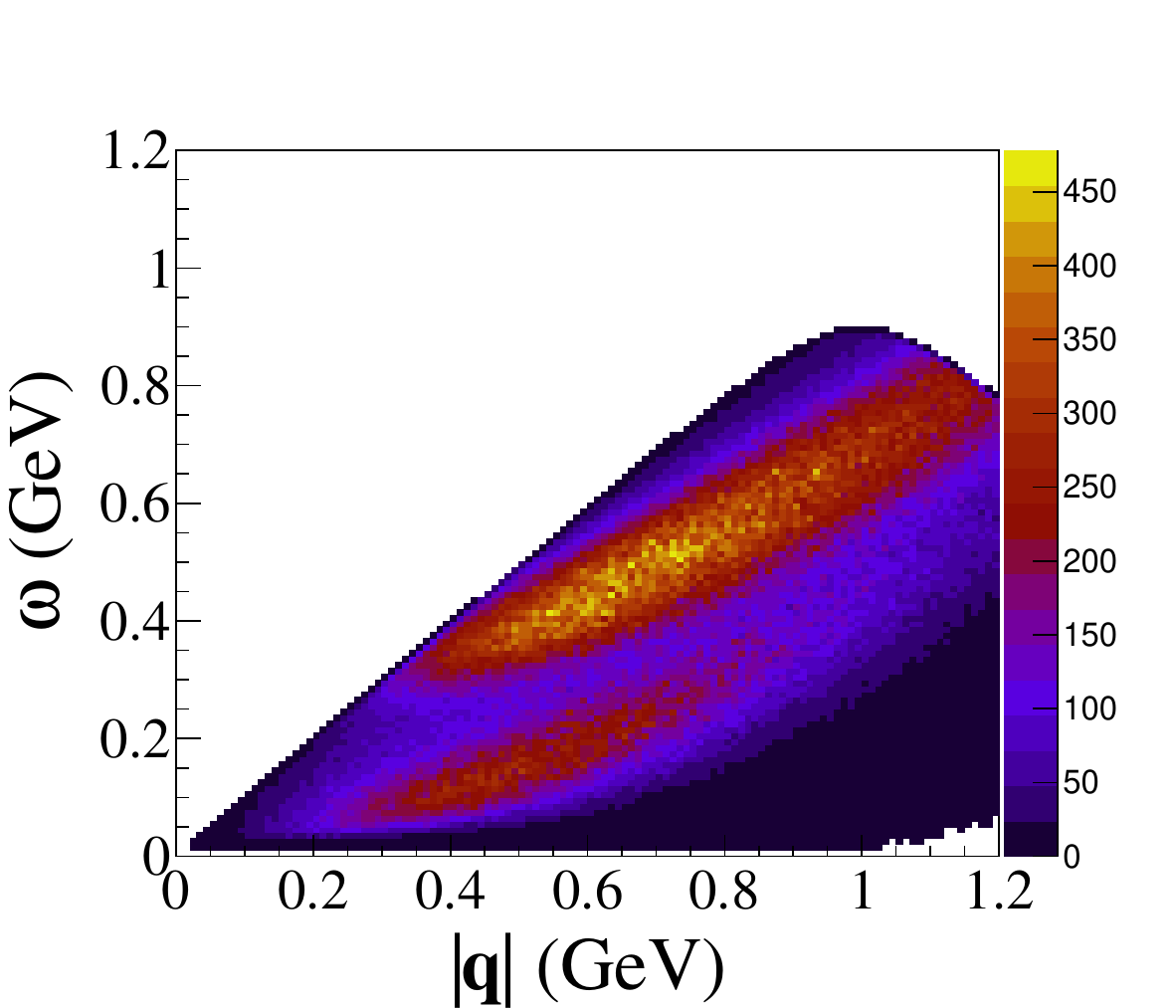}
        \end{adjustbox}
    \end{minipage}
        \begin{minipage}[b]{0.7\linewidth}
        \begin{adjustbox}{width=\linewidth}
            \includegraphics[width=\linewidth]{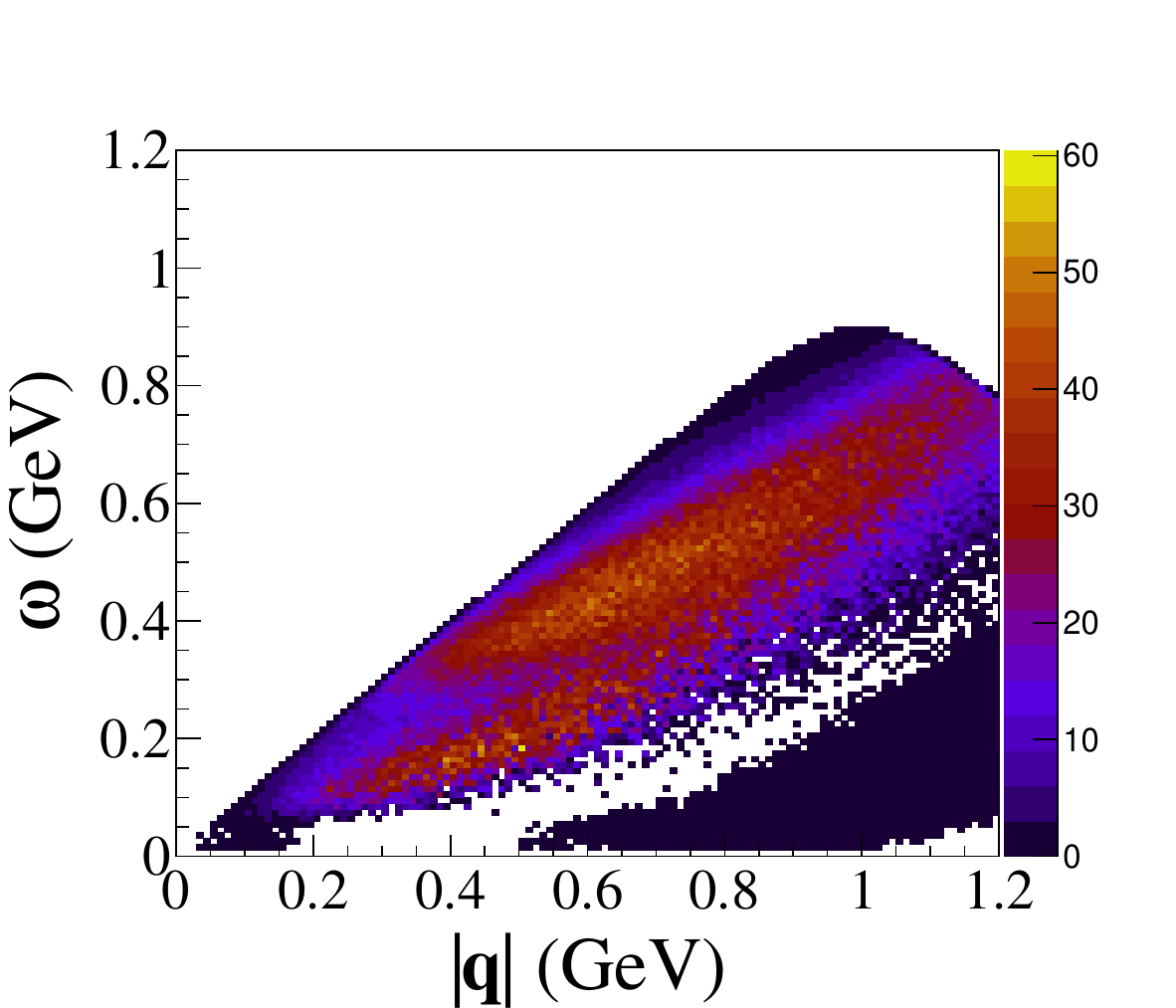}
        \end{adjustbox}
    \end{minipage}
        \begin{minipage}[b]{0.7\linewidth}
        \begin{adjustbox}{width=\linewidth}
            \includegraphics[width=\linewidth]{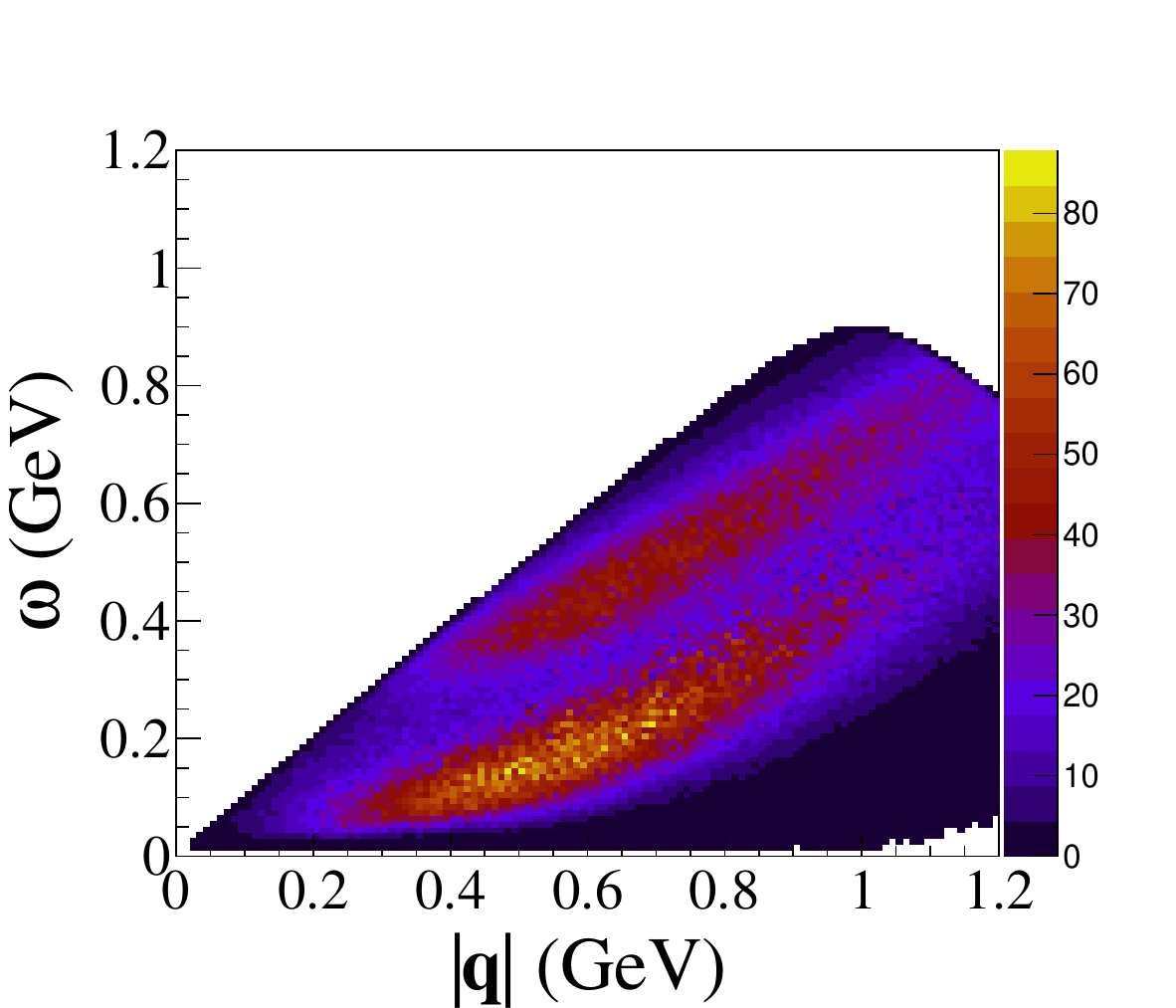}
        \end{adjustbox}
    \end{minipage}
    \caption{\label{fig:inc_phasespace_C12} Double differential cross section $d^2\sigma/dq d\omega$ produced by 2020 Valencia model for $pp$ (\textbf{top}), $np$ (\textbf{middle}), and $pn$ (\textbf{bottom}) outgoing pair in \textit{2p2h} mechanism. }
\end{figure}

The development of the exclusive MEC model by the Valencia group (see Ref.~\cite{Sobczyk:2020dkn}), referred to as the 2020 Valencia model, separates \textit{2p2h} and \textit{3p3h} contributions to the total MEC cross section. In addition, the 2020 Valencia model further decomposes the \textit{2p2h} contribution into three parts based on the isospin of the outgoing nucleon pair namely $pp$, $np$, and $pn$. The model also predicts the momenta of the outgoing nucleon states in all three cases. The correlation between the two nucleons is modeled differently for different nucleon pairs, see Fig.~\ref{fig:inc_phasespace_C12}. 

\begin{figure*}
    \centering
    \includegraphics[width=0.7\linewidth]{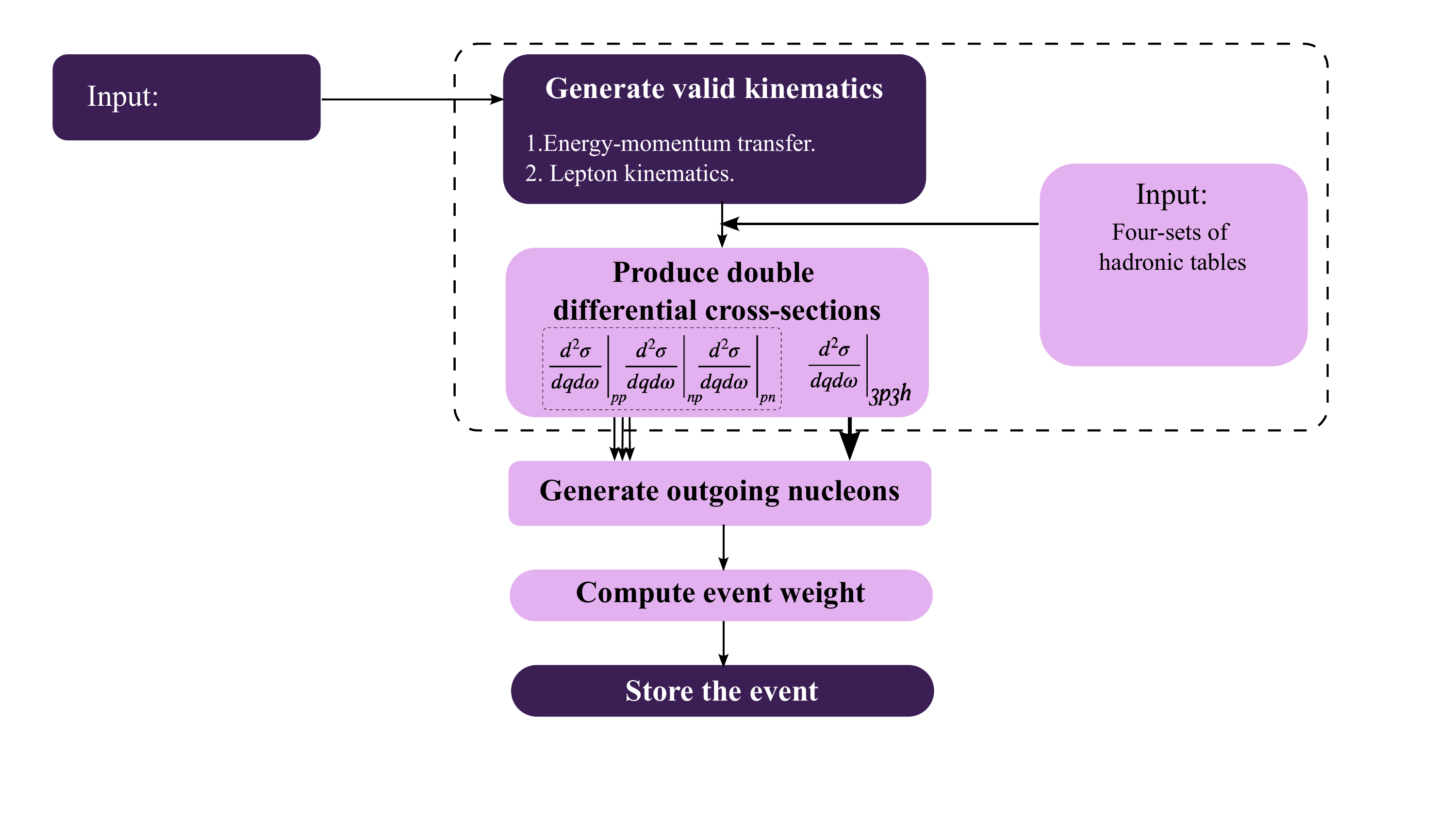}
    \caption{\label{fig:inclusive_part} Algorithm of the new MC MEC model in the context of NuWro implementation of 2020 Valencia model. The ``boxed'' region is the \textit{inclusive part} of the MEC model.}
\end{figure*}

\begin{figure*}
    \centering
    \includegraphics[width=0.7\linewidth]{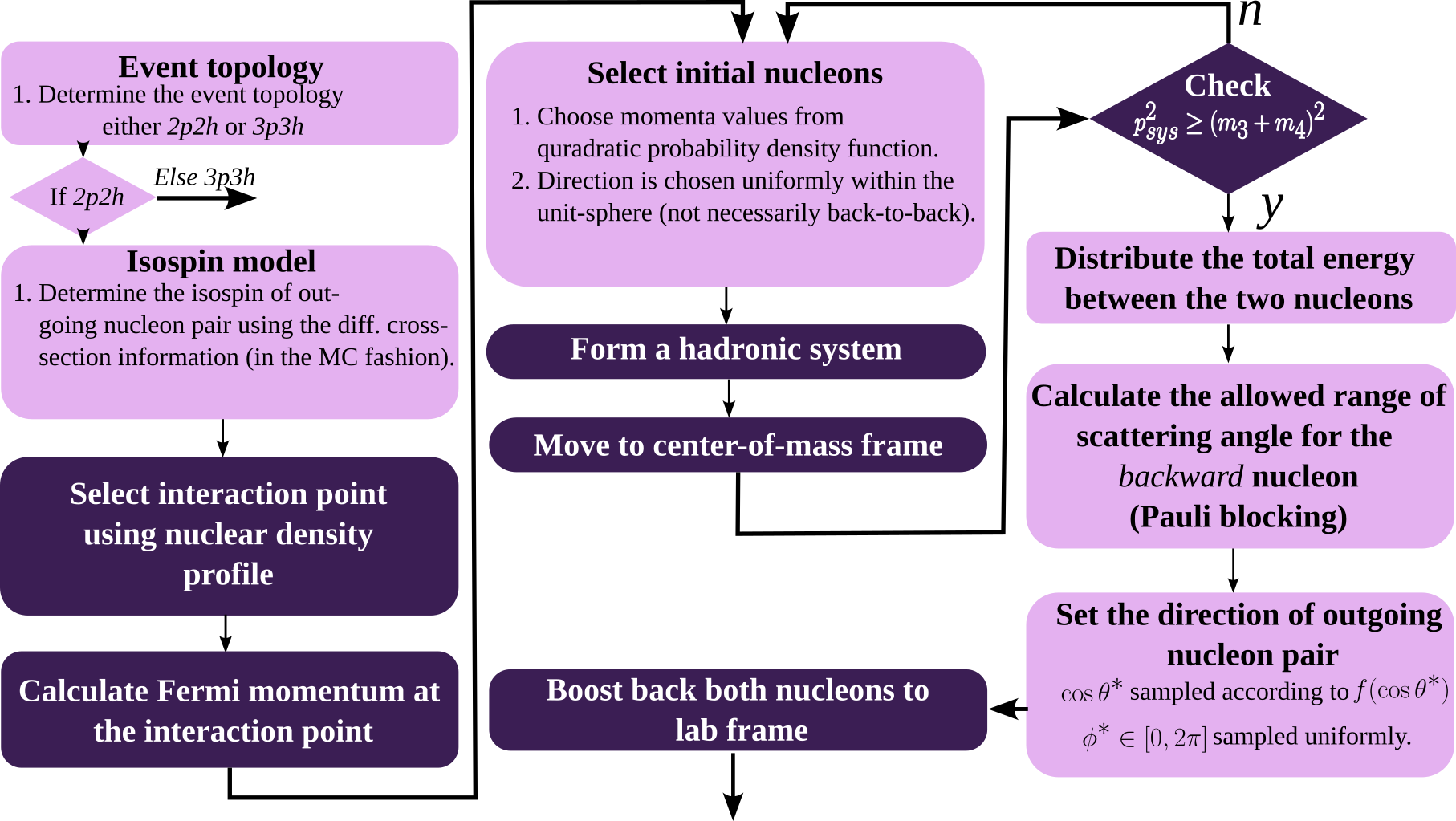}
    \caption{\label{fig:exclusive_part}Algorithm corresponding to ``Generate outgoing nucleons'' given in Fig.~\ref{fig:inclusive_part} in the case of \textit{2p2h} mechanism. The  explicit form of $f(\cos\theta^*)$ is given in Eq.(B1) Appendix B, of Ref.~\cite{prasad2024newmultinucleonknockoutmodel}.}
\end{figure*}

The upcoming version of {\sc NuWro} will include the implementation of the 2020 Valencia model. The algorithm for this implementation is illustrated in Fig.~\ref{fig:inclusive_part} and \ref{fig:exclusive_part}. {\sc NuWro} implementation tries to approximate the correlations between the outgoing nucleon states using a nucleon sampling function \cite{prasad2024newmultinucleonknockoutmodel}. This function has two adjustable parameters that produce a nucleon phase space similar to that of the 2020 Valencia model. 
Here we present the algorithm of the new MC MEC model~\cite{prasad2024newmultinucleonknockoutmodel} in the context of NuWro implementation of 2020 Valencia model. 

The impact of exclusive observables from experimental analyses with the MINER\ensuremath{\nu}A form factor and the new MEC model implemented within {\sc NuWro} is demonstrated in Ref.~\cite{prasad2024newmultinucleonknockoutmodel}. {\sc NuWro} with the new hadronic model when compared with MINER\ensuremath{\nu}A CC$1p0\pi$ data, has a significant impact on all the exclusive variables and produces lower $\chi^2/$d.o.f except in the case of $|\mathbf{p}_n|$ (reconstructed neutron momentum). The implementation of exclusive models in Monte Carlo generators is essential for newer long-baseline neutrino oscillation experiments, like Hyper-Kamiokande~\cite{Hyper-KamiokandeProto-:2015xww} and DUNE~\cite{DUNE:2016hlj}, which require highly accurate modeling of neutrino-nucleus interactions to meet their performance demands~\cite{NuSTEC:2017hzk}.

\subsection{\label{subsec:RES} Ghent hybrid Model}

 In a neutrino-nucleus interaction, pion production occurs via two primary mechanisms: resonant and non-resonant production. Currently, {\sc NuWro} implementation utilizes a dedicated $\Delta (1232)$ resonance production and decay model for the resonant contribution. This model accurately describes the inclusive cross section for pion production through $\Delta (1232)$ resonance formation ~\cite{PhysRevD.80.093001} but the approach becomes less reliable in the higher invariant mass ($W$) region due to the absence of explicit contributions from other resonances. This limitation is overcome by incorporating \verb+PYTHIA+~\cite{tbr} within {\sc NuWro} for modeling the final states
originating from other resonances within the framework of non-resonant production, using inclusive cross sections from the Bodek-Yang approach ~\cite{bodek2003modelingneutrinoelectronscattering}. At low $W$, the contribution from resonant and non-resonant production in the $\Delta$ region is computed via the linear interpolation of the two. The use of \verb+PYTHIA+ hadronization models at low W and low squared-four momentum $(Q^2)$ limits the model performance.\\

The new ``hybrid model''~\cite{yan2024ghenthybridmodelnuwro} from the Ghent group developed for single-pion production (SPP) targets distinct kinematic regions based on $W$ differently and removes the \verb+PYTHIA+ contribution to SPP. The description of SPP is further improved by adding higher resonance contributions, $P_{33}(1232)(\Delta), D_{13}(1520), S_{11}(1535), P_{11}(1440)$, in the second resonance region, interference with the non-resonant background~\cite{PhysRevD.76.033005}, and a Regge description at high $W$. The transition from the low energy regime to the high energy regime is described in ~\cite{yan2024ghenthybridmodelnuwro}.
 

The model demonstrates improved agreement in comparisons to experimental data from MINER\ensuremath{\nu}A and T2K measurements, particularly in pion kinematics and transverse kinematic imbalance. However, the challenges remain, notably in modeling nuclear effects, which motivates future enhancements to the initial state and interaction mechanisms within this new framework.

\section{\label{sec:ML_techniques}MACHINE LEARNING TECHNIQUES}
We will now discuss the recent advancements made by the Wroc{\l}aw Neutrino Group in machine learning techniques, specifically regarding lepton-nucleus interactions. The aim was to explore model-independent reconstructions of such interactions and to parameterize cross sections across different nuclear species.

In this context, we employed a deep neural network (DNN) framework to obtain empirical fits for the cross sections for electron scattering on carbon over a broad kinematic range. extending from the quasi-elastic peak, resonance excitation, up to the onset of deep inelastic scattering~\cite{PhysRevC.110.025501}. 

We developed two types of models. The first model was based on a bootstrap approach, while the second relies on a Monte Carlo dropout technique.  To evaluate the performance of these models, we compared their predictions against test datasets that were excluded from the training process, as well as against theoretical predictions derived from the spectral function approach. In Ref.~\cite{PhysRevC.110.025501}, (see Figs. 6 and 7 there) we presented comparisons of the predictions of the two models to six different datasets. The bootstrap model performed better than the MC dropout model and accurately reproduced the data at both low and high scattering angles.\\

The extent of DNN was further explored by investigating transfer learning (TL), a technique  that allows deep neural networks to adapt to new problems with limited data (see Ref.~\cite{graczyk2024electronnucleuscrosssectionstransfer}). A pre-trained DNN model, developed using electron-carbon scattering data (bootstrap model), was used to predict cross sections for electron interactions with nuclear targets such as lithium, oxygen, aluminum, calcium, and iron. 



Our conclusion was that DNNs trained on carbon data can effectively predict cross sections for various nuclear targets, even with limited measurements in relevant kinematic regions with proper fine-tuning.\\


\begin{acknowledgments}
This work was supported by the National Science Center under the Grant No.UMO-2021/41/B/ST2/02778. K.M.G is also partly supported by the {\sl Excellence Initiative – Research University,  2020-2026} at
the University of Wroclaw.
\end{acknowledgments}

\bibliography{main}

\providecommand{\noopsort}[1]{}\providecommand{\singleletter}[1]{#1}%
\begin{thebibliography}{22}%
\makeatletter
\providecommand \@ifxundefined [1]{%
 \@ifx{#1\undefined}
}%
\providecommand \@ifnum [1]{%
 \ifnum #1\expandafter \@firstoftwo
 \else \expandafter \@secondoftwo
 \fi
}%
\providecommand \@ifx [1]{%
 \ifx #1\expandafter \@firstoftwo
 \else \expandafter \@secondoftwo
 \fi
}%
\providecommand \natexlab [1]{#1}%
\providecommand \enquote  [1]{``#1''}%
\providecommand \bibnamefont  [1]{#1}%
\providecommand \bibfnamefont [1]{#1}%
\providecommand \citenamefont [1]{#1}%
\providecommand \href@noop [0]{\@secondoftwo}%
\providecommand \href [0]{\begingroup \@sanitize@url \@href}%
\providecommand \@href[1]{\@@startlink{#1}\@@href}%
\providecommand \@@href[1]{\endgroup#1\@@endlink}%
\providecommand \@sanitize@url [0]{\catcode `\\12\catcode `\$12\catcode `\&12\catcode `\#12\catcode `\^12\catcode `\_12\catcode `\%12\relax}%
\providecommand \@@startlink[1]{}%
\providecommand \@@endlink[0]{}%
\providecommand \url  [0]{\begingroup\@sanitize@url \@url }%
\providecommand \@url [1]{\endgroup\@href {#1}{\urlprefix }}%
\providecommand \urlprefix  [0]{URL }%
\providecommand \Eprint [0]{\href }%
\providecommand \doibase [0]{https://doi.org/}%
\providecommand \selectlanguage [0]{\@gobble}%
\providecommand \bibinfo  [0]{\@secondoftwo}%
\providecommand \bibfield  [0]{\@secondoftwo}%
\providecommand \translation [1]{[#1]}%
\providecommand \BibitemOpen [0]{}%
\providecommand \bibitemStop [0]{}%
\providecommand \bibitemNoStop [0]{.\EOS\space}%
\providecommand \EOS [0]{\spacefactor3000\relax}%
\providecommand \BibitemShut  [1]{\csname bibitem#1\endcsname}%
\let\auto@bib@innerbib\@empty
\bibitem [{\citenamefont {Juszczak}\ \emph {et~al.}(2006)\citenamefont {Juszczak}, \citenamefont {Nowak},\ and\ \citenamefont {Sobczyk}}]{Juszczak:2005zs}%
  \BibitemOpen
  \bibfield  {author} {\bibinfo {author} {\bibfnamefont {C.}~\bibnamefont {Juszczak}}, \bibinfo {author} {\bibfnamefont {J.~A.}\ \bibnamefont {Nowak}},\ and\ \bibinfo {author} {\bibfnamefont {J.~T.}\ \bibnamefont {Sobczyk}},\ }\bibfield  {title} {\bibinfo {title} {{Simulations from a new neutrino event generator}},\ }\href {https://doi.org/10.1016/j.nuclphysbps.2006.08.069} {\bibfield  {journal} {\bibinfo  {journal} {Nucl. Phys. B Proc. Suppl.}\ }\textbf {\bibinfo {volume} {159}},\ \bibinfo {pages} {211} (\bibinfo {year} {2006})},\ \Eprint {https://arxiv.org/abs/hep-ph/0512365} {arXiv:hep-ph/0512365} \BibitemShut {NoStop}%
\bibitem [{\citenamefont {Golan}\ \emph {et~al.}(2012)\citenamefont {Golan}, \citenamefont {Juszczak},\ and\ \citenamefont {Sobczyk}}]{Golan:2012wx}%
  \BibitemOpen
  \bibfield  {author} {\bibinfo {author} {\bibfnamefont {T.}~\bibnamefont {Golan}}, \bibinfo {author} {\bibfnamefont {C.}~\bibnamefont {Juszczak}},\ and\ \bibinfo {author} {\bibfnamefont {J.~T.}\ \bibnamefont {Sobczyk}},\ }\bibfield  {title} {\bibinfo {title} {{Final State Interactions Effects in Neutrino-Nucleus Interactions}},\ }\href {https://doi.org/10.1103/PhysRevC.86.015505} {\bibfield  {journal} {\bibinfo  {journal} {Phys. Rev. C}\ }\textbf {\bibinfo {volume} {86}},\ \bibinfo {pages} {015505} (\bibinfo {year} {2012})},\ \Eprint {https://arxiv.org/abs/1202.4197} {arXiv:1202.4197 [nucl-th]} \BibitemShut {NoStop}%
\bibitem [{\citenamefont {Banerjee}\ \emph {et~al.}(2024)\citenamefont {Banerjee}, \citenamefont {Ankowski}, \citenamefont {Graczyk}, \citenamefont {Kowal}, \citenamefont {Prasad},\ and\ \citenamefont {Sobczyk}}]{Banerjee:2023hub}%
  \BibitemOpen
  \bibfield  {author} {\bibinfo {author} {\bibfnamefont {R.~D.}\ \bibnamefont {Banerjee}}, \bibinfo {author} {\bibfnamefont {A.~M.}\ \bibnamefont {Ankowski}}, \bibinfo {author} {\bibfnamefont {K.~M.}\ \bibnamefont {Graczyk}}, \bibinfo {author} {\bibfnamefont {B.~E.}\ \bibnamefont {Kowal}}, \bibinfo {author} {\bibfnamefont {H.}~\bibnamefont {Prasad}},\ and\ \bibinfo {author} {\bibfnamefont {J.~T.}\ \bibnamefont {Sobczyk}},\ }\bibfield  {title} {\bibinfo {title} {{JLab spectral functions of argon in nuwro and their implications for MicroBooNE}},\ }\href {https://doi.org/10.1103/PhysRevD.109.073004} {\bibfield  {journal} {\bibinfo  {journal} {Phys. Rev. D}\ }\textbf {\bibinfo {volume} {109}},\ \bibinfo {pages} {073004} (\bibinfo {year} {2024})},\ \Eprint {https://arxiv.org/abs/2312.13369} {arXiv:2312.13369 [hep-ph]} \BibitemShut {NoStop}%
\bibitem [{\citenamefont {Jiang}\ \emph {et~al.}(2022)\citenamefont {Jiang}, \citenamefont {Ankowski}, \citenamefont {Abrams} \emph {et~al.}}]{PhysRevD.105.112002}%
  \BibitemOpen
  \bibfield  {author} {\bibinfo {author} {\bibfnamefont {L.}~\bibnamefont {Jiang}}, \bibinfo {author} {\bibfnamefont {A.~M.}\ \bibnamefont {Ankowski}}, \bibinfo {author} {\bibfnamefont {D.}~\bibnamefont {Abrams}}, \emph {et~al.} (\bibinfo {collaboration} {Jefferson Lab Hall A Collaboration}),\ }\bibfield  {title} {\bibinfo {title} {Determination of the argon spectral function from $(e,{e}^{\ensuremath{'}}p)$ data},\ }\href {https://doi.org/10.1103/PhysRevD.105.112002} {\bibfield  {journal} {\bibinfo  {journal} {Phys. Rev. D}\ }\textbf {\bibinfo {volume} {105}},\ \bibinfo {pages} {112002} (\bibinfo {year} {2022})}\BibitemShut {NoStop}%
\bibitem [{\citenamefont {Jiang}\ \emph {et~al.}(2023)\citenamefont {Jiang}, \citenamefont {Ankowski}, \citenamefont {Abrams} \emph {et~al.}}]{PhysRevD.107.012005}%
  \BibitemOpen
  \bibfield  {author} {\bibinfo {author} {\bibfnamefont {L.}~\bibnamefont {Jiang}}, \bibinfo {author} {\bibfnamefont {A.~M.}\ \bibnamefont {Ankowski}}, \bibinfo {author} {\bibfnamefont {D.}~\bibnamefont {Abrams}}, \emph {et~al.} (\bibinfo {collaboration} {The Jefferson Lab Hall A Collaboration}),\ }\bibfield  {title} {\bibinfo {title} {Determination of the titanium spectral function from $(e,\text{ }{e}^{\ensuremath{'}}p)$ data},\ }\href {https://doi.org/10.1103/PhysRevD.107.012005} {\bibfield  {journal} {\bibinfo  {journal} {Phys. Rev. D}\ }\textbf {\bibinfo {volume} {107}},\ \bibinfo {pages} {012005} (\bibinfo {year} {2023})}\BibitemShut {NoStop}%
\bibitem [{\citenamefont {Ankowski}\ \emph {et~al.}(2015)\citenamefont {Ankowski}, \citenamefont {Benhar},\ and\ \citenamefont {Sakuda}}]{PhysRevD.91.033005}%
  \BibitemOpen
  \bibfield  {author} {\bibinfo {author} {\bibfnamefont {A.~M.}\ \bibnamefont {Ankowski}}, \bibinfo {author} {\bibfnamefont {O.}~\bibnamefont {Benhar}},\ and\ \bibinfo {author} {\bibfnamefont {M.}~\bibnamefont {Sakuda}},\ }\bibfield  {title} {\bibinfo {title} {Improving the accuracy of neutrino energy reconstruction in charged-current quasielastic scattering off nuclear targets},\ }\href {https://doi.org/10.1103/PhysRevD.91.033005} {\bibfield  {journal} {\bibinfo  {journal} {Phys. Rev. D}\ }\textbf {\bibinfo {volume} {91}},\ \bibinfo {pages} {033005} (\bibinfo {year} {2015})}\BibitemShut {NoStop}%
\bibitem [{\citenamefont {Abratenko}\ \emph {et~al.}(2024)\citenamefont {Abratenko}, \citenamefont {Acciarri}, \citenamefont {Adams} \emph {et~al.}}]{2024}%
  \BibitemOpen
  \bibfield  {author} {\bibinfo {author} {\bibfnamefont {P.}~\bibnamefont {Abratenko}}, \bibinfo {author} {\bibfnamefont {R.}~\bibnamefont {Acciarri}}, \bibinfo {author} {\bibfnamefont {C.}~\bibnamefont {Adams}}, \emph {et~al.},\ }\bibfield  {title} {\bibinfo {title} {Scintillation light in sbnd: simulation, reconstruction, and expected performance of the photon detection system},\ }\bibfield  {journal} {\bibinfo  {journal} {The European Physical Journal C}\ }\textbf {\bibinfo {volume} {84}},\ \href {https://doi.org/10.1140/epjc/s10052-024-13306-3} {10.1140/epjc/s10052-024-13306-3} (\bibinfo {year} {2024})\BibitemShut {NoStop}%
\bibitem [{\citenamefont {Aguilar}\ \emph {et~al.}(2001)\citenamefont {Aguilar}, \citenamefont {Auerbach}, \citenamefont {Burman} \emph {et~al.}}]{PhysRevD.64.112007}%
  \BibitemOpen
  \bibfield  {author} {\bibinfo {author} {\bibfnamefont {A.}~\bibnamefont {Aguilar}}, \bibinfo {author} {\bibfnamefont {L.~B.}\ \bibnamefont {Auerbach}}, \bibinfo {author} {\bibfnamefont {R.~L.}\ \bibnamefont {Burman}}, \emph {et~al.} (\bibinfo {collaboration} {LSND Collaboration}),\ }\bibfield  {title} {\bibinfo {title} {Evidence for neutrino oscillations from the observation of ${\overline{\ensuremath{\nu}}}_{e}$ appearance in a ${\overline{\ensuremath{\nu}}}_{\ensuremath{\mu}}$ beam},\ }\href {https://doi.org/10.1103/PhysRevD.64.112007} {\bibfield  {journal} {\bibinfo  {journal} {Phys. Rev. D}\ }\textbf {\bibinfo {volume} {64}},\ \bibinfo {pages} {112007} (\bibinfo {year} {2001})}\BibitemShut {NoStop}%
\bibitem [{\citenamefont {Abratenko}\ \emph {et~al.}(2020)\citenamefont {Abratenko}, \citenamefont {Alrashed}, \citenamefont {An} \emph {et~al.}}]{PhysRevLett.125.201803}%
  \BibitemOpen
  \bibfield  {author} {\bibinfo {author} {\bibfnamefont {P.}~\bibnamefont {Abratenko}}, \bibinfo {author} {\bibfnamefont {M.}~\bibnamefont {Alrashed}}, \bibinfo {author} {\bibfnamefont {R.}~\bibnamefont {An}}, \emph {et~al.} (\bibinfo {collaboration} {MicroBooNE Collaboration}),\ }\bibfield  {title} {\bibinfo {title} {First measurement of differential charged current quasielasticlike ${\ensuremath{\nu}}_{\ensuremath{\mu}}$-argon scattering cross sections with the microboone detector},\ }\href {https://doi.org/10.1103/PhysRevLett.125.201803} {\bibfield  {journal} {\bibinfo  {journal} {Phys. Rev. Lett.}\ }\textbf {\bibinfo {volume} {125}},\ \bibinfo {pages} {201803} (\bibinfo {year} {2020})}\BibitemShut {NoStop}%
\bibitem [{\citenamefont {Sobczyk}(2012)}]{Sobczyk:2012ms}%
  \BibitemOpen
  \bibfield  {author} {\bibinfo {author} {\bibfnamefont {J.~T.}\ \bibnamefont {Sobczyk}},\ }\bibfield  {title} {\bibinfo {title} {{Multinucleon Ejection Model for Meson Exchange Current Neutrino Interactions}},\ }\href {https://doi.org/10.1103/PhysRevC.86.015504} {\bibfield  {journal} {\bibinfo  {journal} {Phys. Rev. C}\ }\textbf {\bibinfo {volume} {86}},\ \bibinfo {pages} {015504} (\bibinfo {year} {2012})},\ \Eprint {https://arxiv.org/abs/1201.3673} {arXiv:1201.3673 [hep-ph]} \BibitemShut {NoStop}%
\bibitem [{\citenamefont {Sobczyk}\ \emph {et~al.}(2020)\citenamefont {Sobczyk}, \citenamefont {Nieves},\ and\ \citenamefont {S\'anchez}}]{Sobczyk:2020dkn}%
  \BibitemOpen
  \bibfield  {author} {\bibinfo {author} {\bibfnamefont {J.~E.}\ \bibnamefont {Sobczyk}}, \bibinfo {author} {\bibfnamefont {J.}~\bibnamefont {Nieves}},\ and\ \bibinfo {author} {\bibfnamefont {F.}~\bibnamefont {S\'anchez}},\ }\bibfield  {title} {\bibinfo {title} {{Exclusive-final-state hadron observables from neutrino-nucleus multinucleon knockout}},\ }\href {https://doi.org/10.1103/PhysRevC.102.024601} {\bibfield  {journal} {\bibinfo  {journal} {Phys. Rev. C}\ }\textbf {\bibinfo {volume} {102}},\ \bibinfo {pages} {024601} (\bibinfo {year} {2020})},\ \Eprint {https://arxiv.org/abs/2002.08302} {arXiv:2002.08302 [nucl-th]} \BibitemShut {NoStop}%
\bibitem [{\citenamefont {Prasad}\ \emph {et~al.}(2024)\citenamefont {Prasad}, \citenamefont {Sobczyk}, \citenamefont {Ankowski}, \citenamefont {Bonilla}, \citenamefont {Banerjee}, \citenamefont {Graczyk},\ and\ \citenamefont {Kowal}}]{prasad2024newmultinucleonknockoutmodel}%
  \BibitemOpen
  \bibfield  {author} {\bibinfo {author} {\bibfnamefont {H.}~\bibnamefont {Prasad}}, \bibinfo {author} {\bibfnamefont {J.~T.}\ \bibnamefont {Sobczyk}}, \bibinfo {author} {\bibfnamefont {A.~M.}\ \bibnamefont {Ankowski}}, \bibinfo {author} {\bibfnamefont {J.~L.}\ \bibnamefont {Bonilla}}, \bibinfo {author} {\bibfnamefont {R.~D.}\ \bibnamefont {Banerjee}}, \bibinfo {author} {\bibfnamefont {K.~M.}\ \bibnamefont {Graczyk}},\ and\ \bibinfo {author} {\bibfnamefont {B.~E.}\ \bibnamefont {Kowal}},\ }\href {https://arxiv.org/abs/2411.11523} {\bibinfo {title} {New multinucleon knockout model in nuwro monte carlo generator}} (\bibinfo {year} {2024}),\ \Eprint {https://arxiv.org/abs/2411.11523} {arXiv:2411.11523 [hep-ph]} \BibitemShut {NoStop}%
\bibitem [{\citenamefont {Abe}\ \emph {et~al.}(2015)\citenamefont {Abe} \emph {et~al.}}]{Hyper-KamiokandeProto-:2015xww}%
  \BibitemOpen
  \bibfield  {author} {\bibinfo {author} {\bibfnamefont {K.}~\bibnamefont {Abe}} \emph {et~al.} (\bibinfo {collaboration} {Hyper-Kamiokande Proto-}),\ }\bibfield  {title} {\bibinfo {title} {{Physics potential of a long-baseline neutrino oscillation experiment using a J-PARC neutrino beam and Hyper-Kamiokande}},\ }\href {https://doi.org/10.1093/ptep/ptv061} {\bibfield  {journal} {\bibinfo  {journal} {PTEP}\ }\textbf {\bibinfo {volume} {2015}},\ \bibinfo {pages} {053C02} (\bibinfo {year} {2015})},\ \Eprint {https://arxiv.org/abs/1502.05199} {arXiv:1502.05199 [hep-ex]} \BibitemShut {NoStop}%
\bibitem [{\citenamefont {Acciarri}\ \emph {et~al.}(2016)\citenamefont {Acciarri} \emph {et~al.}}]{DUNE:2016hlj}%
  \BibitemOpen
  \bibfield  {author} {\bibinfo {author} {\bibfnamefont {R.}~\bibnamefont {Acciarri}} \emph {et~al.} (\bibinfo {collaboration} {DUNE}),\ }\bibfield  {title} {\bibinfo {title} {{Long-Baseline Neutrino Facility (LBNF) and Deep Underground Neutrino Experiment (DUNE)}: {Conceptual Design Report, Volume 1: The LBNF and DUNE Projects}},\ }\href@noop {} {\  (\bibinfo {year} {2016})},\ \Eprint {https://arxiv.org/abs/1601.05471} {arXiv:1601.05471 [physics.ins-det]} \BibitemShut {NoStop}%
\bibitem [{\citenamefont {Alvarez-Ruso}\ \emph {et~al.}(2018)\citenamefont {Alvarez-Ruso} \emph {et~al.}}]{NuSTEC:2017hzk}%
  \BibitemOpen
  \bibfield  {author} {\bibinfo {author} {\bibfnamefont {L.}~\bibnamefont {Alvarez-Ruso}} \emph {et~al.} (\bibinfo {collaboration} {NuSTEC}),\ }\bibfield  {title} {\bibinfo {title} {{NuSTEC White Paper: Status and challenges of neutrino\textendash{}nucleus scattering}},\ }\href {https://doi.org/10.1016/j.ppnp.2018.01.006} {\bibfield  {journal} {\bibinfo  {journal} {Prog. Part. Nucl. Phys.}\ }\textbf {\bibinfo {volume} {100}},\ \bibinfo {pages} {1} (\bibinfo {year} {2018})},\ \Eprint {https://arxiv.org/abs/1706.03621} {arXiv:1706.03621 [hep-ph]} \BibitemShut {NoStop}%
\bibitem [{\citenamefont {Graczyk}\ \emph {et~al.}(2009)\citenamefont {Graczyk}, \citenamefont {Kie\l{}czewska}, \citenamefont {Przew\l{}ocki},\ and\ \citenamefont {Sobczyk}}]{PhysRevD.80.093001}%
  \BibitemOpen
  \bibfield  {author} {\bibinfo {author} {\bibfnamefont {K.~M.}\ \bibnamefont {Graczyk}}, \bibinfo {author} {\bibfnamefont {D.}~\bibnamefont {Kie\l{}czewska}}, \bibinfo {author} {\bibfnamefont {P.}~\bibnamefont {Przew\l{}ocki}},\ and\ \bibinfo {author} {\bibfnamefont {J.~T.}\ \bibnamefont {Sobczyk}},\ }\bibfield  {title} {\bibinfo {title} {${C}_{5}^{A}$ axial form factor from bubble chamber experiments},\ }\href {https://doi.org/10.1103/PhysRevD.80.093001} {\bibfield  {journal} {\bibinfo  {journal} {Phys. Rev. D}\ }\textbf {\bibinfo {volume} {80}},\ \bibinfo {pages} {093001} (\bibinfo {year} {2009})}\BibitemShut {NoStop}%
\bibitem [{\citenamefont {Sjöstrand}\ \emph {et~al.}(2006)\citenamefont {Sjöstrand}, \citenamefont {Mrenna},\ and\ \citenamefont {Skands}}]{tbr}%
  \BibitemOpen
  \bibfield  {author} {\bibinfo {author} {\bibfnamefont {T.}~\bibnamefont {Sjöstrand}}, \bibinfo {author} {\bibfnamefont {S.}~\bibnamefont {Mrenna}},\ and\ \bibinfo {author} {\bibfnamefont {P.}~\bibnamefont {Skands}},\ }\bibfield  {title} {\bibinfo {title} {Pythia 6.4 physics and manual},\ }\href {https://doi.org/10.1088/1126-6708/2006/05/026} {\bibfield  {journal} {\bibinfo  {journal} {Journal of High Energy Physics}\ }\textbf {\bibinfo {volume} {2006}},\ \bibinfo {pages} {026} (\bibinfo {year} {2006})}\BibitemShut {NoStop}%
\bibitem [{\citenamefont {Bodek}\ and\ \citenamefont {Yang}(2003)}]{bodek2003modelingneutrinoelectronscattering}%
  \BibitemOpen
  \bibfield  {author} {\bibinfo {author} {\bibfnamefont {A.}~\bibnamefont {Bodek}}\ and\ \bibinfo {author} {\bibfnamefont {U.~K.}\ \bibnamefont {Yang}},\ }\href {https://arxiv.org/abs/hep-ex/0308007} {\bibinfo {title} {Modeling neutrino and electron scattering inelastic cross sections}} (\bibinfo {year} {2003}),\ \Eprint {https://arxiv.org/abs/hep-ex/0308007} {arXiv:hep-ex/0308007 [hep-ex]} \BibitemShut {NoStop}%
\bibitem [{\citenamefont {Yan}\ \emph {et~al.}(2024)\citenamefont {Yan}, \citenamefont {Niewczas}, \citenamefont {Nikolakopoulos}, \citenamefont {González-Jiménez}, \citenamefont {Jachowicz}, \citenamefont {Lu}, \citenamefont {Sobczyk},\ and\ \citenamefont {Zheng}}]{yan2024ghenthybridmodelnuwro}%
  \BibitemOpen
  \bibfield  {author} {\bibinfo {author} {\bibfnamefont {Q.}~\bibnamefont {Yan}}, \bibinfo {author} {\bibfnamefont {K.}~\bibnamefont {Niewczas}}, \bibinfo {author} {\bibfnamefont {A.}~\bibnamefont {Nikolakopoulos}}, \bibinfo {author} {\bibfnamefont {R.}~\bibnamefont {González-Jiménez}}, \bibinfo {author} {\bibfnamefont {N.}~\bibnamefont {Jachowicz}}, \bibinfo {author} {\bibfnamefont {X.}~\bibnamefont {Lu}}, \bibinfo {author} {\bibfnamefont {J.}~\bibnamefont {Sobczyk}},\ and\ \bibinfo {author} {\bibfnamefont {Y.}~\bibnamefont {Zheng}},\ }\href {https://arxiv.org/abs/2405.05212} {\bibinfo {title} {The ghent hybrid model in nuwro: a new neutrino single-pion production model in the gev regime}} (\bibinfo {year} {2024}),\ \Eprint {https://arxiv.org/abs/2405.05212} {arXiv:2405.05212 [hep-ph]} \BibitemShut {NoStop}%
\bibitem [{\citenamefont {Hern\'andez}\ \emph {et~al.}(2007)\citenamefont {Hern\'andez}, \citenamefont {Nieves},\ and\ \citenamefont {Valverde}}]{PhysRevD.76.033005}%
  \BibitemOpen
  \bibfield  {author} {\bibinfo {author} {\bibfnamefont {E.}~\bibnamefont {Hern\'andez}}, \bibinfo {author} {\bibfnamefont {J.}~\bibnamefont {Nieves}},\ and\ \bibinfo {author} {\bibfnamefont {M.}~\bibnamefont {Valverde}},\ }\bibfield  {title} {\bibinfo {title} {Weak pion production off the nucleon},\ }\href {https://doi.org/10.1103/PhysRevD.76.033005} {\bibfield  {journal} {\bibinfo  {journal} {Phys. Rev. D}\ }\textbf {\bibinfo {volume} {76}},\ \bibinfo {pages} {033005} (\bibinfo {year} {2007})}\BibitemShut {NoStop}%
\bibitem [{\citenamefont {Kowal}\ \emph {et~al.}(2024)\citenamefont {Kowal}, \citenamefont {Graczyk}, \citenamefont {Ankowski}, \citenamefont {Banerjee}, \citenamefont {Prasad},\ and\ \citenamefont {Sobczyk}}]{PhysRevC.110.025501}%
  \BibitemOpen
  \bibfield  {author} {\bibinfo {author} {\bibfnamefont {B.~E.}\ \bibnamefont {Kowal}}, \bibinfo {author} {\bibfnamefont {K.~M.}\ \bibnamefont {Graczyk}}, \bibinfo {author} {\bibfnamefont {A.~M.}\ \bibnamefont {Ankowski}}, \bibinfo {author} {\bibfnamefont {R.~D.}\ \bibnamefont {Banerjee}}, \bibinfo {author} {\bibfnamefont {H.}~\bibnamefont {Prasad}},\ and\ \bibinfo {author} {\bibfnamefont {J.~T.}\ \bibnamefont {Sobczyk}},\ }\bibfield  {title} {\bibinfo {title} {Empirical fits to inclusive electron-carbon scattering data obtained by deep-learning methods},\ }\href {https://doi.org/10.1103/PhysRevC.110.025501} {\bibfield  {journal} {\bibinfo  {journal} {Phys. Rev. C}\ }\textbf {\bibinfo {volume} {110}},\ \bibinfo {pages} {025501} (\bibinfo {year} {2024})}\BibitemShut {NoStop}%
\bibitem [{\citenamefont {Graczyk}\ \emph {et~al.}(2024)\citenamefont {Graczyk}, \citenamefont {Kowal}, \citenamefont {Ankowski}, \citenamefont {Banerjee}, \citenamefont {Bonilla}, \citenamefont {Prasad},\ and\ \citenamefont {Sobczyk}}]{graczyk2024electronnucleuscrosssectionstransfer}%
  \BibitemOpen
  \bibfield  {author} {\bibinfo {author} {\bibfnamefont {K.~M.}\ \bibnamefont {Graczyk}}, \bibinfo {author} {\bibfnamefont {B.~E.}\ \bibnamefont {Kowal}}, \bibinfo {author} {\bibfnamefont {A.~M.}\ \bibnamefont {Ankowski}}, \bibinfo {author} {\bibfnamefont {R.~D.}\ \bibnamefont {Banerjee}}, \bibinfo {author} {\bibfnamefont {J.~L.}\ \bibnamefont {Bonilla}}, \bibinfo {author} {\bibfnamefont {H.}~\bibnamefont {Prasad}},\ and\ \bibinfo {author} {\bibfnamefont {J.~T.}\ \bibnamefont {Sobczyk}},\ }\href {https://arxiv.org/abs/2408.09936} {\bibinfo {title} {Electron-nucleus cross sections from transfer learning}} (\bibinfo {year} {2024}),\ \Eprint {https://arxiv.org/abs/2408.09936} {arXiv:2408.09936 [hep-ph]} \BibitemShut {NoStop}%
\end{thebibliography}%

\end{document}